\documentclass[a4paper,10pt,twoside]{cpc-hepnp}
\usepackage{CJK,upgreek,fancyhdr}
\usepackage{multicol}
\usepackage{graphicx}
\usepackage{booktabs}
\usepackage{amssymb,bm,mathrsfs,bbm,amscd}
\usepackage[tbtags]{amsmath}
\usepackage{lastpage}
\usepackage{cases}
\usepackage{subfigure}
\usepackage{epstopdf}
\usepackage{bm}

\usepackage[colorlinks,linkcolor=black,anchorcolor=black,citecolor=black,urlcolor=black]{hyperref}

\newcommand{\jpsi}{$J/\psi$~}
\newcommand{\jpsiwos}{$J/\psi$}
\newcommand{\psip}{$\psi(3686)$~}

\newcommand{\eetoee}{$e^+e^- \to e^+e^-$~}
\newcommand{\eetomumu}{$e^+e^- \to \mu^+\mu^-$~}

\newcommand{\eetoeewos}{$e^+e^- \to e^+e^-$}
\newcommand{\eetomumuwos}{$e^+e^- \to \mu^+\mu^-$}

\newcommand{\Tw}{\Gamma_{\rm tot}}
\newcommand{\Ew}{\Gamma_{ee}}
\newcommand{\Mw}{\Gamma_{\mu\mu}}

\newcommand{\csth}{\cos\theta}

\begin{document}
\begin{CJK*}{UTF8}{gbsn}

\fancyhead[c]{\small Accepted by Chinese Physics C}
\fancyfoot[C]{\small xxxxxx-\thepage}
\footnotetext[0]{}

\title{Analytic Forms for Cross Sections of Di-lepton Production \\ from $e^+e^-$ Collisions around the \jpsi Resonance \thanks{Supported by National Natural Science Foundation of China (11275211) and Istituto Nazionale di Fisica Nucleare, Italy}}

\author{Xing-Yu Zhou (周兴玉)$^{1;1)}$ \email{xyzhou@ihep.ac.cn} \quad Ya-Di Wang (王雅迪)$^{2;2)}$ \email{Y.Wang@him.uni-mainz.de} \quad Li-Gang Xia (夏力钢)$^{3;3)}$ \email{xialigang@tsinghua.edu.cn}}

\maketitle

\address
{
	$^1$ Institute of High Energy Physics, Chinese Academy of Sciences, Beijing 100049, China \\
	$^2$ Helmholtz Institute Mainz, Mainz 55128, Germany \\
	$^3$ Department of Physics, Tsinghua University, Beijing 100084, China
}

\begin{abstract}
	A detailed theoretical derivation of the cross sections of \eetoee and \eetomumu around the \jpsi resonance is reported. The resonance and interference parts of the cross sections, related to \jpsi resonance parameters, are calculated. Higher-order corrections for vacuum polarization and initial-state radiation are considered. An arbitrary upper limit of radiative correction integration is involved. Full and simplified versions of analytic formulae are given with precision at the level of 0.1\% and 0.2\%, respectively. Moreover, the results obtained in the paper can be applied to the case of the \psip resonance.
\end{abstract}



\footnotetext[0]{\ \\}

\begin{multicols}{2}

\section{Introduction}
The \jpsi resonance is frequently referred to as a hydrogen atom for QCD, and its resonance parameters (mass $M$, total width $\Tw$, leptonic widths $\Ew$ and $\Mw$, and so on) describe the fundamental properties of the strong and electromagnetic interactions. In theory, the decay widths can be predicted by different potential \\ models \cite{potential models 1,potential models 2} and lattice QCD calculations \cite{lattice calculations}. In experiment, with results from BABAR \cite{BABAR}, CLEO \cite{CLEO} and KEDR \cite{KEDR}, determinations of these decay widths have entered a period of precision measurement.

In 2012, data samples were taken at 15 center-of-mass energy points around the \jpsi resonance with the BESIII detector \cite{BESIII detector} operated at the BEPCII collider \cite{BESIII detector}. In this energy region, BEPCII provides high luminosity and BESIII shows excellent performance, which helps us accurately measure  the cross sections of \eetoee and \eetomumuwos. To measure \jpsi decay widths, accurate theoretical formulae taking into account higher-order corrections are also needed. If one wishes to have a high-efficiency optimization procedure, it is better to have analytic expressions for the theoretical cross sections. Because the continuum parts of these cross sections do not involve \jpsi decay widths and can be evaluated precisely by Monte-Carlo generators such as the Babayaga generator \cite{BABAYAGA}, only the analytic forms for the resonance and interference parts are derived in this paper.

We will start with theoretical fundamentals on the structure function method, its applications to the cases of \eetoee and \eetomumuwos, Born cross sections and the vacuum polarization function in Section \ref{Theoretical fundamentals}. Then, we will give the definitions and resulting formulae for the resonance and interference parts of the cross sections of \eetoee and \eetomumu in Section \ref{Calculations of the resonance and interference parts}. Most of the purely mathematical derivation is given in Appendix A to make the text easier to read.

\section{Theoretical fundamentals}
\label{Theoretical fundamentals}

\subsection{Structure function method}

Generally, initial-state radiation (ISR), final-state radiation (FSR) and their interference (ISR-FSR relation) must be considered when one makes higher-order corrections to cross sections. Here, the ISR-FSR relation includes interference of diagrams with emission of real and virtual photons between initial- and final-state particles. The suppression level of the ISR-FSR relation between the production and decay stages of heavy unstable particles is discussed in Ref. \cite{ISR-FSR relation}. According to the conclusion in Ref. \cite{ISR-FSR relation}, there is no need to take into account the ISR-FSR relation in the case of \jpsi, because it is suppressed by $\Gamma_{\rm tot}/M$ (about $3\times10^{-5}$). As for FSR, a universal calculation is impossible if one has no explicit knowledge of selection criteria, so it needs to be handled separately with a numerical method, which is outside the scope of this paper. Thus, in this paper the calculation with ISR only is presented.

The structure function method \cite{STRUCTUREFUNCTIONMETHOD} is adopted here to deal with ISR. Its fundamental formula is
\begin{equation}
	\sigma(s) = \int \int_0^X \frac{d\bar{\sigma}}{d\Omega}(s(1-x),\cos\theta) F(s,x) dx d\Omega.
	\label{Equation: Fundamental formula of structure function method}
\end{equation}
Here, $\sigma$ stands for the cross section after correction, $\frac{d\bar{\sigma}}{d\Omega}$ for the differential cross section before correction, $F$ for the radiator, $s$ for the square of the center-of-mass energy and $\theta$ for the polar angle of the positively charged final particle in the center-of-mass frame. The upper limit $X$ of the integration variable $x$ is usually set as $1-s^{\prime}_{min}/s$, where $s^{\prime}_{min}$ is the minimum of the invariant mass squared of the final-state particle system excluding the emitted photons.

The radiator $F$ adopted in this paper was first derived in Ref. \cite{Radiator 1} and slightly revised in Ref. \cite{Radiator 2}. Both documents are in Chinese, although the former has an English-language preprint (Ref. \cite{Radiator 3}). It is different from but a very good approximation of the classical one in Ref. \cite{STRUCTUREFUNCTIONMETHOD}. Its expression is
\begin{align}
	F(s,x) & = x^{v-1}v(1+\delta) \nonumber \\
	& + x^{v}\left(-v-\frac{v^2}{4}\right) + x^{v+1}\left(\frac{v}{2}-\frac{3}{8}v^2\right),
	\label{Equation: Expression of radiator.}
\end{align}
where
\begin{equation}
	\delta(v) = \frac{\alpha}{\pi}(\frac{\pi^2}{3}-\frac{1}{2})+\frac{3}{4}v+\left(\frac{9}{32}-\frac{\pi^2}{12}\right)v^2
\end{equation}
and
\begin{equation}
	v(s) = \frac{2\alpha}{\pi}\left(\ln\frac{s}{m_e^2}-1\right).
\end{equation}
Here, $\alpha$ stands for the fine structure constant and $m_e$ denotes the electron mass.

\subsection{Applications of the structure fuction meth-\\od to \eetoee and \eetomumu}
\label{Subsection: Applications of the structure fuction method to eetoee and eetomumu}

Applying the structure function method to the cases of \eetoee and \eetomumuwos, one can get
\begin{align}
	&\ \ \  \left(\frac{d\sigma}{d\Omega}\right)_{ee|\mu\mu}(s,\cos\theta) \nonumber \\
	& = \int_0^X \left(\frac{d\bar{\sigma}}{d\Omega}\right)_{ee|\mu\mu}(s(1-x),\cos\theta) F(s,x) dx,
	\label{Equation: Formula of structure function method applied to eetoee and eetomumu.}
\end{align}
where the symbol $|$ stands for ``or",
\begin{align}
	\left(\frac{d\bar{\sigma}}{d\Omega}\right)_{ee} & = \left(\frac{d\sigma_0}{d\Omega}\right)_{ee}^{\rm S}\left|\frac{1}{1-\Pi(s)}\right|^2 \nonumber \\
	& + \left(\frac{d\sigma_0}{d\Omega}\right)_{ee}^{\rm T}\left|\frac{1}{1-\Pi(t)}\right|^2 \nonumber \\
	& + \left(\frac{d\sigma_0}{d\Omega}\right)_{ee}^{\rm STI} Re\left(\frac{1}{1-\Pi(s)}\overline{\frac{1}{1-\Pi(t)}}\right) \label{Equation: Cross sections of eetoee with vacuum polarization considered}
\end{align}
and
\begin{equation}
	\left(\frac{d\bar{\sigma}}{d\Omega}\right)_{\mu\mu} = \left(\frac{d\sigma_0}{d\Omega}\right)_{\mu\mu}^{\rm S}\left|\frac{1}{1-\Pi(s)}\right|^2. \label{Equation: Cross sections of eetomumu with vacuum polarization considered}
\end{equation}
Here, $t$ denotes the square of the 4-momentum transferred in the t channel. As for \eetoeewos, the relation between $t$ and $s$ is
\begin{equation}
	t \approx -\frac{s}{2}(1-\csth).
\end{equation}
In addition, $\left(\frac{d\sigma_0}{d\Omega}\right)_{ee}^{\rm S}$, $\left(\frac{d\sigma_0}{d\Omega}\right)_{ee}^{\rm T}$, $\left(\frac{d\sigma_0}{d\Omega}\right)_{ee}^{\rm STI}$ and $\left(\frac{d\sigma_0}{d\Omega}\right)_{\mu\mu}^{\rm S}$ are Born cross sections, and $\frac{1}{1-\Pi}$ is the vacuum polarization function. They will be discussed in the following two subsections.

\subsection{Born cross sections}

The quantities $\left(\frac{d\sigma_0}{d\Omega}\right)_{ee}^{\rm S}$, $\left(\frac{d\sigma_0}{d\Omega}\right)_{ee}^{\rm T}$ and $\left(\frac{d\sigma_0}{d\Omega}\right)_{ee}^{\rm STI}$ are the s channel part, the t channel part and the s-t interference part of the Born cross section of \eetoee $\left(\left(\frac{d\sigma_0}{d\Omega}\right)_{ee}\right)$, respectively, that is
\begin{align}
	& \ \ \  \left(\frac{d\sigma_0}{d\Omega}\right)_{ee} = \left(\frac{d\sigma_0}{d\Omega}\right)_{ee}^{\rm S} + \left(\frac{d\sigma_0}{d\Omega}\right)_{ee}^{\rm T} + \left(\frac{d\sigma_0}{d\Omega}\right)_{ee}^{\rm STI},
\end{align}
where
\begin{subequations}
	\label{Align: Born cross section of eetoee}
	\begin{align}
		\left(\frac{d\sigma_0}{d\Omega}\right)_{ee}^{\rm S} & = \frac{\alpha^2}{4s} (1+\cos^2\theta), \label{Equation: S channel part of Born cross section of eetoee} \\
		\left(\frac{d\sigma_0}{d\Omega}\right)_{ee}^{\rm T} & = \frac{\alpha^2}{2s} \frac{(1+\cos\theta)^2+4}{(1-\cos\theta)^2}, \label{Equation: T channel part of Born cross section of eetoee} \\
		\left(\frac{d\sigma_0}{d\Omega}\right)_{ee}^{\rm STI} & = - \frac{\alpha^2}{2s} \frac{(1+\cos\theta)^2}{1-\cos\theta}. \label{Equation: S-T interference part of Born cross section of eetoee}
	\end{align}
\end{subequations}
The Born cross section of \eetomumu $\left(\left(\frac{d\sigma_0}{d\Omega}\right)_{\mu\mu}\right)$ has only an s channel part $\left(\frac{d\sigma_0}{d\Omega}\right)_{\mu\mu}^{\rm S}$, which equals exactly $\left(\frac{d\sigma_0}{d\Omega}\right)_{ee}^{\rm S}$ given by Eq. (\ref{Equation: S channel part of Born cross section of eetoee}).

\subsection{Vacuum polarization function}

In Section 4 of Ref. \cite{KEDRpsip}, the distinction and relationship between the ``bare" and ``dressed" parameters of $J^{PC}=1^{--}$ resonances (for example \jpsiwos) are discussed in detail. In the discussion there, the vacuum polarization function is written as
\begin{equation}
	\frac{1}{1-\Pi(q^2)} = \frac{1}{1-\Pi_{\rm 0}(q^2)}+\Pi_{\rm R}(q^2),
	\label{Equation: The vacuum polarization function}
\end{equation}
where $\Pi_{\rm R}$ is expressed with the ``dressed" parameters $M$, $\Gamma_{\rm tot}$ and $\Gamma_{ee}$ as
\begin{equation}
	\Pi_{\rm R}(q^2) = \frac{3\Gamma_{ee}}{\alpha} \frac{q^2}{M} \frac{1}{q^2-M^2+iM\Gamma_{\rm tot}}.
	\label{Equation: Resonance part of vacuum polarization function}
\end{equation}
Here, $\Pi_{\rm R}$ stands for the contribution from the resonance itself (in our case, it is \jpsiwos), while $\Pi_{\rm 0}$ denotes contributions from other sources. Based on the lepton universality assumption, $\Gamma_{ee}$ in Eq. (\ref{Equation: Resonance part of vacuum polarization function}) can be substituted by $\sqrt{\Gamma_{ee}\Gamma_{\mu\mu}}$ in the case of \eetomumuwos.

According to Eq. (\ref{Equation: The vacuum polarization function}), $\frac{1}{1-\Pi(s)}$ and $\frac{1}{1-\Pi(t)}$ in Eq. (\ref{Equation: Cross sections of eetoee with vacuum polarization considered}) and (\ref{Equation: Cross sections of eetomumu with vacuum polarization considered}) can be expressed as
\begin{equation}
	\frac{1}{1-\Pi(s)} = \frac{1}{1-\Pi_0(s)} + \Pi_{\rm R}(s) \label{Equation: Vacuum polarization in the timelike region}
\end{equation}
and
\begin{equation}
	\frac{1}{1-\Pi(t)} = \frac{1}{1-\Pi_0(t)}. \label{Equation: Vacuum polarization in the spacelike region}
\end{equation}
No $\Pi_{\rm R}(t)$ term appears in Eq. (\ref{Equation: Vacuum polarization in the spacelike region}) because it can be safely ignored in the spacelike region. Besides, the imaginary parts of $\frac{1}{1-\Pi_0(s)}$ and $\frac{1}{1-\Pi_0(t)}$ can be safely ignored as well. Consequently, $\frac{1}{1-\Pi_0(s)}$ and $\frac{1}{1-\Pi_0(t)}$ will be regarded as real in the following section.

\section{Calculations of the resonance and interference parts}
\label{Calculations of the resonance and interference parts}
\subsection{Definitions}

Considering $\left(\frac{d\bar{\sigma}}{d\Omega}\right)_{ee}$ and $\left(\frac{d\bar{\sigma}}{d\Omega}\right)_{\mu\mu}$ given by Eq. (\ref{Equation: Cross sections of eetoee with vacuum polarization considered}) and (\ref{Equation: Cross sections of eetomumu with vacuum polarization considered}) as well as $\frac{1}{1-\Pi(s)}$ and $\frac{1}{1-\Pi(t)}$ given by Eq. (\ref{Equation: Vacuum polarization in the timelike region}) and (\ref{Equation: Vacuum polarization in the spacelike region}), one can expand $\left(\frac{d\sigma}{d\Omega}\right)_{ee}$ and $\left(\frac{d\sigma}{d\Omega}\right)_{\mu\mu}$ via Eq. (\ref{Equation: Formula of structure function method applied to eetoee and eetomumu.}) into \\ many small terms. With these small terms regrouped, the resonance and interference parts of $\left(\frac{d\sigma}{d\Omega}\right)_{ee}$ and $\left(\frac{d\sigma}{d\Omega}\right)_{\mu\mu}$, namely $\left(\frac{d\sigma}{d\Omega}\right)_{ee}^{\rm R}$, $\left(\frac{d\sigma}{d\Omega}\right)_{ee}^{\rm CRI}$, $\left(\frac{d\sigma}{d\Omega}\right)_{\mu\mu}^{\rm R}$ and $\left(\frac{d\sigma}{d\Omega}\right)_{\mu\mu}^{\rm CRI}$, can be defined as

\end{multicols}

\begin{subequations}
	\label{Align: definitions}
	\begin{align}
		\left(\frac{d\sigma}{d\Omega}\right)_{ee}^{\rm R} & = \int_0^X \left(\frac{d\sigma_0}{d\Omega}\right)_{ee}^{\rm S}(s(1-x),\cos\theta)\left|\Pi_{\rm R}(s(1-x))\right|^2 F(s,x)dx, \label{Equation: Resonance part of cross section of eetoee} \\
		\nonumber \\
		\left(\frac{d\sigma}{d\Omega}\right)_{ee}^{\rm CRI} & = \int_0^X \Bigg(\left(\frac{d\sigma_0}{d\Omega}\right)_{ee}^{\rm S}(s(1-x),\cos\theta) 2Re\left(\frac{1}{1-\Pi_0(s(1-x))}\Pi_{\rm R}(s(1-x))\right) + \nonumber \\
		& \ \ \ \ \ \ \ \ \ \ \  \left(\frac{d\sigma_0}{d\Omega}\right)_{ee}^{\rm STI}(s(1-x),\cos\theta) Re\left(\Pi_{\rm R}(s(1-x))\frac{1}{1-\Pi_0(t(1-x))}\right)\Bigg)F(s,x)dx, \label{Equation: Continuous-Resonance interference part of cross section of eetoee} \\
		\nonumber \\
		\left(\frac{d\sigma}{d\Omega}\right)_{\mu\mu}^{\rm R} & = \int_0^X \left(\frac{d\sigma_0}{d\Omega}\right)_{\mu\mu}^{\rm S}(s(1-x),\cos\theta)\left|\Pi_{\rm R}(s(1-x))\right|^2 F(s,x)dx, \label{Equation: Resonance part of cross section of eetomumu} \\
		\nonumber \\
		\left(\frac{d\sigma}{d\Omega}\right)_{\mu\mu}^{\rm CRI} & = \int_0^X \left(\frac{d\sigma_0}{d\Omega}\right)_{\mu\mu}^{\rm S}(s(1-x),\cos\theta) 2Re\left(\frac{1}{1-\Pi_0(s(1-x))}\Pi_{\rm R}(s(1-x))\right) F(s,x)dx. \label{Equation: Continuous-Resonance interference part of cross section of eetomumu}
	\end{align}
\end{subequations}

\centerline{\rule{80mm}{0.1pt}}

\begin{multicols}{2}
	
	With $\left(\frac{d\sigma_0}{d\Omega}\right)_{ee|\mu\mu}^{\rm S}$ and $\left(\frac{d\sigma_0}{d\Omega}\right)_{ee}^{\rm STI}$ expressed in Eq. (\ref{Equation: S channel part of Born cross section of eetoee}) and (\ref{Equation: S-T interference part of Born cross section of eetoee}) as well as $\Pi_{\rm R}$ expressed in Eq. (\ref{Equation: Resonance part of vacuum polarization function}) further employed, one can rewrite $\left(\frac{d\sigma}{d\Omega}\right)_{ee}^{\rm R}$, $\left(\frac{d\sigma}{d\Omega}\right)_{ee}^{\rm CRI}$, $\left(\frac{d\sigma}{d\Omega}\right)_{\mu\mu}^{\rm R}$ and $\left(\frac{d\sigma}{d\Omega}\right)_{\mu\mu}^{\rm CRI}$ more explicitly as
	
\end{multicols}

\begin{subequations}
	\label{Align: semi-finished results}
	\begin{align}
		\left(\frac{d\sigma}{d\Omega}\right)_{ee}^{\rm R} & = \frac{9\Gamma_{ee}^2}{4M^2} \cdot I^{\rm R} \cdot (1+\cos^2\theta), \label{Equation: Resonance part of cross section of eetoee --- semi-finished results} \\
		\nonumber \\
		\left(\frac{d\sigma}{d\Omega}\right)_{ee}^{\rm CRI} & = \frac{3\Gamma_{ee}\alpha}{2M} \cdot I^{\rm CRI} \cdot \left( (1+\cos^2\theta) \frac{1}{1-\Pi_0(s)} - \frac{(1+\cos\theta)^2}{1-\cos\theta} \frac{1}{1-\Pi_0(t)} \right), \label{Equation: Continuous-Resonance interference part of cross section of eetoee --- semi-finished results} \\
		\nonumber \\
		\left(\frac{d\sigma}{d\Omega}\right)_{\mu\mu}^{\rm R} & = \frac{9\Gamma_{ee}\Gamma_{\mu\mu}}{4M^2} \cdot I^{\rm R} \cdot (1+\cos^2\theta), \label{Equation: Resonance part of cross section of eetomumu --- semi-finished results}
	\end{align}
	\begin{align}
		\left(\frac{d\sigma}{d\Omega}\right)_{\mu\mu}^{\rm CRI} & = \frac{3\sqrt{\Gamma_{ee}\Gamma_{\mu\mu}}\alpha}{2M} \cdot I^{\rm CRI} \cdot (1+\cos^2\theta) \frac{1}{1-\Pi_0(s)} , \label{Equation: Continuous-Resonance interference part of cross section of eetomumu --- semi-finished results}
	\end{align}
\end{subequations}
where
\begin{subequations}
	\label{Align: definitions of the two integrals}
	\begin{align}
		I^{\rm R} & = \int_0^X \frac{s(1-x)}{(s(1-x)-M^2)^2+M^2\Gamma_{\rm tot}^2} F(s,x)dx, \\
		\nonumber \\
		I^{\rm CRI} & = \int_0^X \frac{s(1-x)-M^2}{(s(1-x)-M^2)^2+M^2\Gamma_{\rm tot}^2} F(s,x)dx.
	\end{align}
\end{subequations}

\centerline{\rule{80mm}{0.1pt}}

\begin{multicols}{2}
	Here, in the cases of $\left(\frac{d\sigma}{d\Omega}\right)_{ee}^{\rm CRI}$ and $\left(\frac{d\sigma}{d\Omega}\right)_{\mu\mu}^{\rm CRI}$, $\frac{1}{1-\Pi_0(s)}$ and $\frac{1}{1-\Pi_0(t)}$ are used as very good approximations to the equivalents of $\frac{1}{1-\Pi_0(s(1-x))}$ and $\frac{1}{1-\Pi_0(t(1-x))}$ after integration in Eq. (\ref{Align: definitions}). Numerical calculation indicates that the resulting deviations are less than 0.01\%.
	
	As can be seen from  Eq. (\ref{Align: semi-finished results}), to evaluate further, only $I^{\rm R}$ and $I^{\rm CRI}$ have to be calculated. Detailed calculations of the two integrals are put in Appendix A, which includes three parts: A.1, A.2, A.3. Their analytic formulae are fully derived in part A.1. Due to complexity, simplified versions of the analytic formulae are further obtained in part A.2. Finally, both versions of the analytic formulae are compared with numerical computing results in part A.3.
	
	Based on those of $I^{\rm R}$ and $I^{\rm CRI}$, we will list directly the full and simplified version of analytic results of $\left(\frac{d\sigma}{d\Omega}\right)_{ee}^{\rm R}$, $\left(\frac{d\sigma}{d\Omega}\right)_{ee}^{\rm CRI}$, $\left(\frac{d\sigma}{d\Omega}\right)_{\mu\mu}^{\rm R}$ and $\left(\frac{d\sigma}{d\Omega}\right)_{\mu\mu}^{\rm CRI}$ and discuss briefly their comparisons with numerical computing results in the following three subsections.
	
	\subsection{Full version of analytic results}
	
	With $I^{\rm R}$ and $I^{\rm CRI}$ expressed in Eq. (\ref{Equation: the first key integral}) and (\ref{Equation: the second key integral}) adopted, the full versions of the analytic formulae for $\left(\frac{d\sigma}{d\Omega}\right)_{ee}^{\rm R}$, $\left(\frac{d\sigma}{d\Omega}\right)_{ee}^{\rm CRI}$, $\left(\frac{d\sigma}{d\Omega}\right)_{\mu\mu}^{\rm R}$ and $\left(\frac{d\sigma}{d\Omega}\right)_{\mu\mu}^{\rm CRI}$ can be written as
	
\end{multicols}

\begin{subequations}
	\label{Align: full version of the final results}
	\begin{align}
		\left(\frac{d\sigma}{d\Omega}\right)_{ee}^{\rm R} & = \frac{9\Gamma_{ee}^2}{4M^2} \cdot s(P - Q) \cdot (1+\cos^2\theta), \\
		\nonumber \\
		\left(\frac{d\sigma}{d\Omega}\right)_{ee}^{\rm CRI} & = \frac{3\Gamma_{ee}\alpha}{2M} \cdot ((s-M^2) P  - s Q ) \cdot \left( (1+\cos^2\theta) \frac{1}{1-\Pi_0(s)} - \frac{(1+\cos\theta)^2}{1-\cos\theta} \frac{1}{1-\Pi_0(t)} \right), \\
		\nonumber \\
		\left(\frac{d\sigma}{d\Omega}\right)_{\mu\mu}^{\rm R} & = \frac{9\Gamma_{ee}\Gamma_{\mu\mu}}{4M^2} \cdot s(P - Q) \cdot  (1+\cos^2\theta), \\
		\nonumber \\
		\left(\frac{d\sigma}{d\Omega}\right)_{\mu\mu}^{\rm CRI} & = \frac{3\sqrt{\Gamma_{ee}\Gamma_{\mu\mu}}\alpha}{2M} \cdot ((s-M^2) P - s Q) \cdot (1+\cos^2\theta) \frac{1}{1-\Pi_0(s)},
	\end{align}
\end{subequations}
where
\begin{subequations}
	\begin{align}
		P & = \frac{1}{s^2}(A\ G(a,\beta,v,X) + B\ G(a,\beta,v+1,X) + C\ H(a,\beta,v,X)), \\
		Q & = \frac{1}{s^2}(D\ G(a,\beta,v+1,X) + E\ H(a,\beta,v,X) + C\ H(a,\beta,v+1,X))
	\end{align}
\end{subequations}
\centerline{\rule{80mm}{0.1pt}}
\begin{multicols}{2}
	\noindent with
	\begin{subequations}
		\begin{align}
			a & = \sqrt{\left(\frac{M^2}{s}-1\right)^2+\frac{M^2\Gamma_{\rm tot}^2}{s^2}},
		\end{align}
		\begin{align}
			\beta & = \cos^{-1}\left(\frac{\left(\frac{M^2}{s}-1\right)}{\sqrt{\left(\frac{M^2}{s}-1\right)^2+\frac{M^2\Gamma_{\rm tot}^2}{s^2}}}\right),
		\end{align}
		\begin{align}
			A & = 1+\delta, \\
			B & = \frac{1}{v+1}\left(-v-\frac{v^2}{4}\right), \\
			C & = \frac{v}{2}-\frac{3}{8}v^2, \\
			D & = \frac{Av}{v+1}, \\
			E & = B(v+1)
		\end{align}
	\end{subequations}
	and
	\begin{subequations}
		\begin{align}
			&\ \ \  G(a,\beta,v,X) \nonumber \\
			& = a^{v-2} \left(\frac{\pi v}{\sin\pi v}\right) \left(\frac{\sin[(1-v)\beta]}{\sin\beta}\right) + v X^{v-4} \bigg(\frac{X^2}{v-2} \nonumber \\
			& +\frac{2a(\cos\beta) X}{v-3}-\frac{a^2(4\cos^2\beta-1)}{v-4}\bigg) \ \ \ (0<v<2),
		\end{align}
		\begin{align}
			&\ \ \  H(a,\beta,v,X) \nonumber \\
			& = h(a\sin\beta,a\cos\beta,v+1,X+a\cos\beta) \nonumber \\
			& - h(a\sin\beta,a\cos\beta,v+1,a\cos\beta), \\
			&\ \ \  h(a,b,c,x) = -\frac{i}{2ac} \nonumber \\
			& \cdot \Bigg(\left(\frac{1}{-ia+x}\right)^{-c}{}_2\mathbb{F}_1\left(-c,-c,1-c,\frac{a+ib}{a+ix}\right) \nonumber \\
			& - \left(\frac{1}{ia+x}\right)^{-c}{}_2\mathbb{F}_1\left(-c,-c,1-c,\frac{ia+b}{ia+x}\right)\Bigg).
		\end{align}
	\end{subequations}
	Here, ${}_2\mathbb{F}_1$ is the Gauss hypergeometric function.
	
	\subsection{Simplified version of analytic results}
	
	With $I^{\rm R}$ and $I^{\rm CRI}$ given by Eq. (\ref{Equation: The approximate results 1}) and (\ref{Equation: The approximate results 2}), the simplified versions of the analytic formulae for $\left(\frac{d\sigma}{d\Omega}\right)_{ee}^{\rm R}$, $\left(\frac{d\sigma}{d\Omega}\right)_{ee}^{\rm CRI}$, $\left(\frac{d\sigma}{d\Omega}\right)_{\mu\mu}^{\rm R}$ and $\left(\frac{d\sigma}{d\Omega}\right)_{\mu\mu}^{\rm CRI}$ can be written as
	
\end{multicols}

\begin{subequations}
	\label{Align: Simplified version of the final results}
	\begin{align}
		\left(\frac{d\sigma}{d\Omega}\right)_{ee}^{\rm R} & = \frac{9\Gamma_{ee}^2}{4M^3\Gamma_{\rm tot}} (1+\delta) Im\mathcal{F} \cdot (1+\cos^2\theta), \label{Equation: The approximate results 1} \\
		\nonumber \\
		\left(\frac{d\sigma}{d\Omega}\right)_{ee}^{\rm CRI} & = - \frac{3\Gamma_{ee}\alpha}{2 M s} (1+\delta) Re\mathcal{F} \cdot \left( (1+\cos^2\theta) \frac{1}{1-\Pi_0(s)} - \frac{(1+\cos\theta)^2}{1-\cos\theta} \frac{1}{1-\Pi_0(t)} \right), \label{Equation: The approximate results 2} \\
		\nonumber \\
		\left(\frac{d\sigma}{d\Omega}\right)_{\mu\mu}^{\rm R} & = \frac{9\Gamma_{ee}\Gamma_{\mu\mu}}{4M^3\Gamma_{\rm tot}} (1+\delta) Im\mathcal{F} \cdot (1+\cos^2\theta), \label{Equation: The approximate results 3} \\
		\nonumber \\
		\left(\frac{d\sigma}{d\Omega}\right)_{\mu\mu}^{\rm CRI} & = - \frac{3\sqrt{\Gamma_{ee}\Gamma_{\mu\mu}}\alpha}{2 M s} (1+\delta) Re\mathcal{F} \cdot (1+\cos^2\theta) \frac{1}{1-\Pi_0(s)}, \label{Equation: The approximate results 4}
	\end{align}
\end{subequations}

\centerline{\rule{80mm}{0.1pt}}

\begin{multicols}{2}
	\noindent where
	\begin{equation}
		\mathcal{F} = \left( \frac{\pi v}{\sin\pi v} \right) \left(\frac{s}{M^2 - s - i M\Gamma_{\rm tot}}\right)^{1-v}.
		\label{Equation: The expressions of mathcalF.}
	\end{equation}
	
	\subsection{Comparison of analytic and numerical computing results}
	\label{Subsection: Comparisons with numerical computing results}
	
	As one can see from Eq. (\ref{Align: semi-finished results}) and (\ref{Align: definitions of the two integrals}),
	\begin{align*}
		& \ \ \  \left(\frac{\Delta \sigma}{\sigma}\right)_{ee|\mu\mu}^{\rm R|CRI}(\text{F}|\text{S},\text{N}) = \frac{\sigma_{ee|\mu\mu}^{\rm R|CRI}(\text{F}|\text{S})-\sigma_{ee|\mu\mu}^{\rm R|CRI}(\text{N})}{\sigma_{ee|\mu\mu}^{\rm R|CRI}(\text{N})} \\
		& = \frac{I^{\rm R|CRI}(\text{F}|\text{S})-I^{\rm R|CRI}(\text{N})}{I^{\rm R|CRI}(\text{N})} = \left(\frac{\Delta I}{I}\right)^{\rm R|CRI}(\text{F}|\text{S},\text{N}).
	\end{align*}
	%
	%
	%
	\noindent Here, the symbols F, S and N stand for the full version of the analytic results, the simplified version of the analytic results and the numerical computing results, respectively.
	
	According to part A.3 (the last part of Appendix A), from $\sqrt{s}=M-10\Tw$ to $\sqrt{s}=M+10\Tw$ with $X$ set at 1 as well as $M$ and $\Tw$ at their PDG values \cite{PDG2016}:
	\begin{equation*}
		\left(\frac{\Delta \sigma}{\sigma}\right)_{ee|\mu\mu}^{\rm R|CRI}(\text{F},\text{N}) = \left(\frac{\Delta I}{I}\right)^{\rm R|CRI}(\text{F},\text{N})<0.01\%
	\end{equation*}
	and
	\begin{equation*}
		\left(\frac{\Delta \sigma}{\sigma}\right)_{ee|\mu\mu}^{\rm R|CRI}(\text{S},\text{N}) = \left(\frac{\Delta I}{I}\right)^{\rm R|CRI}(\text{S},\text{N})<0.1\%.
	\end{equation*}
	Taking into account the precision of the structure function method itself is 0.1\% \cite{STRUCTUREFUNCTIONMETHOD}, we regard 0.1\% and 0.2\% as the precision of the full and simplified versions of the analytic formulae for $\left(\frac{d\sigma}{d\Omega}\right)_{ee}^{\rm R}$, $\left(\frac{d\sigma}{d\Omega}\right)_{ee}^{\rm CRI}$, $\left(\frac{d\sigma}{d\Omega}\right)_{\mu\mu}^{\rm R}$ and $\left(\frac{d\sigma}{d\Omega}\right)_{\mu\mu}^{\rm CRI}$, respectively.

\section{Conclusions}
We have derived the detailed formulae for the resonance and interference parts of the cross sections of \eetoee and \eetomumu around the \jpsi resonance with higher-order corrections for vacuum polarization and initial-state radiation considered. In the derivation, the arbitrary upper limit of radiative correction integration $X$ has been involved. Two (full and simplified) versions of the analytic formulae are given with precision at the levels of 0.1\% and 0.2\%, which are accurate enough for the measurement of \jpsi decay widths at present.

In our derivation, only a very few steps rely on the values of \jpsi resonance parameters and they can be easily verified to be workable for the case of the \psip resonance. In the coming round of data-taking at BESIII, there is a plan for an energy scan around the \psip resonance for the measurement of the resonance parameters. By that time, the results obtained in this paper will be good references.

\ \\

\acknowledgments{The authors would like to thank Prof. Wei-Guo Li for his suggestion on the contributing as well as Prof. Ping Wang, Prof. Hai-Ming Hu and Prof. Chang-Zheng Yuan for their kind help and beneficial discussions.}

\end{multicols}

\vspace{15mm}
\begin{small}
	\renewcommand{\theequation}{A\arabic{equation}}
	\setcounter{equation}{0}
	\begin{multicols}{2}
		\subsection*{Appendix A}
		\label{Section: Appendix A}
		\noindent{\bf Calculations of $I^{\rm R}$ and $I^{\rm CRI}$} \\
		
		\noindent \textbf{A.1 Full versions of analytic formulae} \\
		
		In the appendix, we evaluate the two integrals $I^{\rm R}$ and $I^{\rm CRI}$ required in Section \ref{Calculations of the resonance and interference parts}. For the convenience of further calculations, it is necessary to make some simple transformations by introducing some new variables. The first transformation is
		\begin{equation}
			\frac{1}{(s(1-x)-M^2)^2+M^2\Gamma_{\rm tot}^2} = \frac{1}{s^2} \frac{1}{x^2+2a(\cos\beta) x+a^2},
		\end{equation}
		where
		\begin{subnumcases}{}
			a = \sqrt{\left(\frac{M^2}{s}-1\right)^2+\frac{M^2\Gamma_{\rm tot}^2}{s^2}}, \\
			\beta = \cos^{-1}\left(\frac{\left(\frac{M^2}{s}-1\right)}{\sqrt{\left(\frac{M^2}{s}-1\right)^2+\frac{M^2\Gamma_{\rm tot}^2}{s^2}}}\right).
		\end{subnumcases}
		
		The second transformation is
		\begin{align}
			F(s,x) & = x^{v-1}v(1+\delta) \nonumber \\
			& + x^{v}\left(-v-\frac{v^2}{4}\right) + x^{v+1}\left(\frac{v}{2}-\frac{3}{8}v^2\right) \nonumber \\
			& = Avx^{v-1} + B(v+1)x^{v} + Cx^{v+1},
		\end{align}
		where
		\begin{subnumcases}{}
			A = 1+\delta, \\
			B = \frac{1}{v+1}\left(-v-\frac{v^2}{4}\right), \\
			C = \frac{v}{2}-\frac{3}{8}v^2.
		\end{subnumcases}
		
		The third transformation is
		\begin{align}
			xF(s,x) & = x^{v}v(1+\delta) \nonumber \\
			& + x^{v+1}\left(-v-\frac{v^2}{4}\right) + x^{v+2}\left(\frac{v}{2}-\frac{3}{8}v^2\right) \nonumber \\
			& = D(v+1)x^{v} + Ex^{v+1} + Cx^{v+2},
		\end{align}
		where
		\begin{subnumcases}{}
			D = \frac{Av}{v+1}, \\
			E = B(v+1).
		\end{subnumcases}
		
		In addition, some integral formulae are crucial for further calculations. From the following two integral formulae
		\begin{align}
			&\ \ \  \int_0^{\infty} \frac{vx^{v-1}}{x^2+2a(\cos\beta) x+a^2} dx \nonumber \\
			& = a^{v-2} \left(\frac{\pi v}{\sin\pi v}\right) \left(\frac{\sin[(1-v)\beta]}{\sin\beta}\right) \ \ \ (0<v<2)
		\end{align}
		and
		\begin{align}
			&\ \ \  \int_{X}^{\infty} \frac{vx^{v-1}}{x^2+2a(\cos\beta) x+a^2} dx \simeq v X^{v-4} \Bigg(-\frac{X^2}{v-2} \nonumber \\
			& - \frac{2a(\cos\beta) X}{v-3}+\frac{a^2(4\cos^2\beta-1)}{v-4}\Bigg) \ \ \ (v<2),
		\end{align}
		one obtains for the first integral formula
		\begin{align}
			&\ \ \  G(a,\beta,v,X) = \int_{0}^{X} \frac{vx^{v-1}}{x^2+2a(\cos\beta) x+a^2} dx \nonumber \\
			& \simeq a^{v-2} \left(\frac{\pi v}{\sin\pi v}\right) \left(\frac{\sin[(1-v)\beta]}{\sin\beta}\right) + v X^{v-4} \bigg(\frac{X^2}{v-2} \nonumber \\
			& +\frac{2a(\cos\beta) X}{v-3}-\frac{a^2(4\cos^2\beta-1)}{v-4}\bigg) \ \ \ (0<v<2).
		\end{align}
		
		The second integral formula is
		\begin{align}
			&\ \ \  H(a,\beta,v,X) = \int_{0}^{X} \frac{x^{v+1}}{x^2+2a(\cos\beta) x+a^2} dx \nonumber \\
			& = \int_{0}^{X} \frac{x^{v+1}}{(x+a\cos\beta)^2+(a\sin\beta)^2} dx \nonumber \\
			& = \int_{a\cos\beta}^{X+a\cos\beta} \frac{(y-a\cos\beta)^{v+1}}{y^2+(a\sin\beta)^2} dy \nonumber \\
			& = h(a\sin\beta,a\cos\beta,v+1,X+a\cos\beta) \nonumber \\
			& - h(a\sin\beta,a\cos\beta,v+1,a\cos\beta),
		\end{align}
		where
		\begin{align}
			&\ \ \  h(a,b,c,x) = \int_0^x \frac{(y-b)^c}{y^2+a^2} dy = -\frac{i}{2ac} \nonumber \\
			& \cdot \Bigg(\left(\frac{1}{-ia+x}\right)^{-c}{}_2\mathbb{F}_1\left(-c,-c,1-c,\frac{a+ib}{a+ix}\right) \nonumber \\
			& - \left(\frac{1}{ia+x}\right)^{-c}{}_2\mathbb{F}_1\left(-c,-c,1-c,\frac{ia+b}{ia+x}\right)\Bigg).
		\end{align}
		Here, ${}_2\mathbb{F}_1$ is the Gauss hypergeometric function.
		
		Using the newly introduced variables and the important integral formulae, we get
	\end{multicols}
	\begin{align}
		P & = \int_0^X \frac{1}{(s(1-x)-M^2)^2+M^2\Gamma_{\rm tot}^2} F(s,x)dx = \frac{1}{s^2} \int_0^X \frac{1}{x^2+2a(\cos\beta) x+a^2} (Avx^{v-1} + B(v+1)x^{v} + Cx^{v+1})dx \nonumber \\
		& = \frac{1}{s^2}\left(A \int_0^X \frac{vx^{v-1}}{x^2+2a(\cos\beta) x+a^2} dx + B \int_0^X \frac{(v+1)x^{v}}{x^2+2a(\cos\beta) x+a^2}dx + C \int_0^X \frac{x^{v+1}}{x^2+2a(\cos\beta) x+a^2} dx\right) \nonumber \\
		& = \frac{1}{s^2}(A\ G(a,\beta,v,X) + B\ G(a,\beta,v+1,X) + C\ H(a,\beta,v,X))
	\end{align}
	and
	\begin{align}
		Q & = \int_0^X \frac{x}{(s(1-x)-M^2)^2+M^2\Gamma_{\rm tot}^2} F(s,x)dx = \frac{1}{s^2} \int_0^X \frac{x}{x^2+2a(\cos\beta) x+a^2} (Avx^{v-1} + B(v+1)x^{v} + Cx^{v+1})dx \nonumber \\
		& = \frac{1}{s^2} \int_0^X \frac{1}{x^2+2a(\cos\beta) x+a^2} (D(v+1)x^{v} + Ex^{v+1} + Cx^{v+2})dx \nonumber \\
		& = \frac{1}{s^2} \left( D \int_0^X \frac{(v+1)x^{v}}{x^2+2a(\cos\beta) x+a^2} dx + E \int_0^X \frac{x^{v+1}}{x^2+2a(\cos\beta) x+a^2} dx + C \int_0^X \frac{x^{v+2}}{x^2+2a(\cos\beta) x+a^2} dx \right) \nonumber \\
		& = \frac{1}{s^2} (D\ G(a,\beta,v+1,X) + E\ H(a,\beta,v,X) + C\ H(a,\beta,v+1,X)),
	\end{align}
	and then get
	\begin{align}
		I^{\rm R} & = \int_0^X \frac{s(1-x)}{(s(1-x)-M^2)^2+M^2\Gamma_{\rm tot}^2} F(s,x)dx = s \int_0^X \frac{1-x}{(s(1-x)-M^2)^2+M^2\Gamma_{\rm tot}^2} F(s,x)dx \nonumber \\
		& = s \left(\int_0^X \frac{1}{(s(1-x)-M^2)^2+M^2\Gamma_{\rm tot}^2} F(s,x)dx - \int_0^X \frac{x}{(s(1-x)-M^2)^2+M^2\Gamma_{\rm tot}^2} F(s,x)dx \right) \nonumber \\
		& = s (P - Q)
		\label{Equation: the first key integral}
	\end{align}
	and
	\begin{align}
		I^{\rm CRI} & = \int_0^X \frac{s(1-x)-M^2}{(s(1-x)-M^2)^2+M^2\Gamma_{\rm tot}^2} F(s,x)dx = \int_0^X \frac{(s-M^2)-s x}{(s(1-x)-M^2)^2+M^2\Gamma_{\rm tot}^2} F(s,x)dx \nonumber \\
		& = (s-M^2) \int_0^X \frac{1}{(s(1-x)-M^2)^2+M^2\Gamma_{\rm tot}^2} F(s,x)dx - s \int_0^X \frac{x}{(s(1-x)-M^2)^2+M^2\Gamma_{\rm tot}^2} F(s,x)dx \nonumber \\
		& =(s-M^2) P - s Q.
		\label{Equation: the second key integral}
	\end{align}
	\centerline{\rule{80mm}{0.1pt}}
	\begin{multicols}{2}
		
		Equations (\ref{Equation: the first key integral}) and (\ref{Equation: the second key integral}) give the analytic formulae for $I^{\rm R}$ and $I^{\rm CRI}$. Since there are no approximations made in the derivation, we refer to the formulae as the full versions of the analytic formulae. Considering all the quantities involved in $P$ and $Q$ ($A$, $B$, $C$ and so on), the results are actually very complicated. For ease of use, simplified versions of the analytic formulae are needed.
		
		\noindent \textbf{A.2 Simplified versions of analytic formulae} \\
		
		In this part, we will make some approximations to obtain simplified versions of the analytic formulae. The first step is to reduce $F(s,x)$ to $x^{v-1}v(1+\delta)$. Since $0 \le x \le 1$ and $v \approx 0.08$ in the \jpsi region, the parts discarded are negligible. This reduction leads to $B=0$, $C=0$, $E=0$.
		
		The second step is to reduce $G(a,\beta,v,X)$ to $a^{v-2} \left(\frac{\pi v}{\sin\pi v}\right) \left(\frac{\sin[(1-v)\beta]}{\sin\beta}\right)$. This reduction means that $X \to +\infty$, which is unreasonable from the physical point of view. However, since $v \approx 0.08$ and $a \in (3\times 10^{-5},\  3\times 10^{-2})$, the reduction itself is a reasonable mathematical approximation when $X$ is large enough. In addition, in the cases of $\left(\frac{d\sigma}{d\Omega}\right)_{ee}^{\rm R}$ and $\left(\frac{d\sigma}{d\Omega}\right)_{\mu\mu}^{\rm R}$, a reasonable reduction of $\sin[(1-v)\beta] - a\sin[(-v)\beta]$ to $\sin[(1-v)\beta]$ is also carried out at this step. With the two steps of approximation applied, one can get
		\begin{equation}
			I^{\rm R} \approx \frac{1}{s a\sin\beta} (1+\delta) a^{v-1} \left(\frac{\pi v}{\sin\pi v}\right) \sin[(1-v)\beta]
		\end{equation}
		and
		\begin{equation}
			I^{\rm CRI} \approx - \frac{1}{s} (1+\delta) a^{v-1} \left(\frac{\pi v}{\sin\pi v}\right) \cos[(1-v)\beta].
		\end{equation}
		
		At this point, if one introduces a complex variable
		\begin{equation}
			\mathcal{F} = \left( \frac{\pi v}{\sin\pi v} \right) (a\cos\beta - i a\sin\beta)^{v-1},
		\end{equation}
		then
		\begin{equation}
			a^{v-1} \left(\frac{\pi v}{\sin\pi v}\right) \sin[(1-v)\beta] = Im\mathcal{F},
			\label{Equation: The Im part of mathcalF}
		\end{equation}
		\begin{equation}
			a^{v-1} \left(\frac{\pi v}{\sin\pi v}\right) \cos[(1-v)\beta] = Re\mathcal{F}.
			\label{Equation: The Re part of mathcalF}
		\end{equation}
		Getting $a$ and $\beta$ back to $\sqrt{\left(\frac{M^2}{s}-1\right)^2+\frac{M^2\Gamma_{\rm tot}^2}{s^2}}$ and $\cos^{-1}\left(\frac{\left(\frac{M^2}{s}-1\right)}{\sqrt{\left(\frac{M^2}{s}-1\right)^2+\frac{M^2\Gamma_{\rm tot}^2}{s^2}}}\right)$, respectively, one has
		\begin{equation}
			a \sin\beta=\frac{M\Gamma_{\rm tot}}{s}
			\label{Equation: ASinBeta}
		\end{equation}
		and
		\begin{equation}
			\mathcal{F} = \left( \frac{\pi v}{\sin\pi v} \right) \left(\frac{s}{M^2 - s - i M\Gamma_{\rm tot}}\right)^{1-v}.
			\label{Equation: The expressions of mathcalF.}
		\end{equation}
		
		With Eq. (\ref{Equation: The Im part of mathcalF}), (\ref{Equation: The Re part of mathcalF}) and (\ref{Equation: ASinBeta}), $I^{\rm R}$ and $I^{\rm CRI}$ can be expressed further as
		\begin{equation}
			I^{\rm R} \approx \frac{1}{M \Gamma_{\rm tot}} (1+\delta) Im\mathcal{F} \label{Equation: The approximate results 1}
		\end{equation}
		and
		\begin{equation}
			I^{\rm CRI} \approx - \frac{1}{s} (1+\delta) Re\mathcal{F}. \label{Equation: The approximate results 2}
		\end{equation}
		These are the simplified versions of the analytic formulae we need. \\
		
		\noindent \textbf{A.3 Comparisons of analytic formulae with numerical computing results} \\
		
		To check the validity of these analytic formulae, we compare them with numerical computing results. In the comparisons, the two integrals $I^{\rm R}$ and $I^{\rm CRI}$ are compared from $\sqrt{s}=M-10\Tw$ to $\sqrt{s}=M+10\Tw$ with $X$ set at 1 as well as $M$ and $\Tw$ at their PDG values \cite{PDG2016}. The results are shown in Fig. \ref{Figure: Comparisons with numerical computing results.}.
	\end{multicols}
	\begin{figure}[!h]
		\centering
		\subfigure{\includegraphics[width=0.4\textwidth]{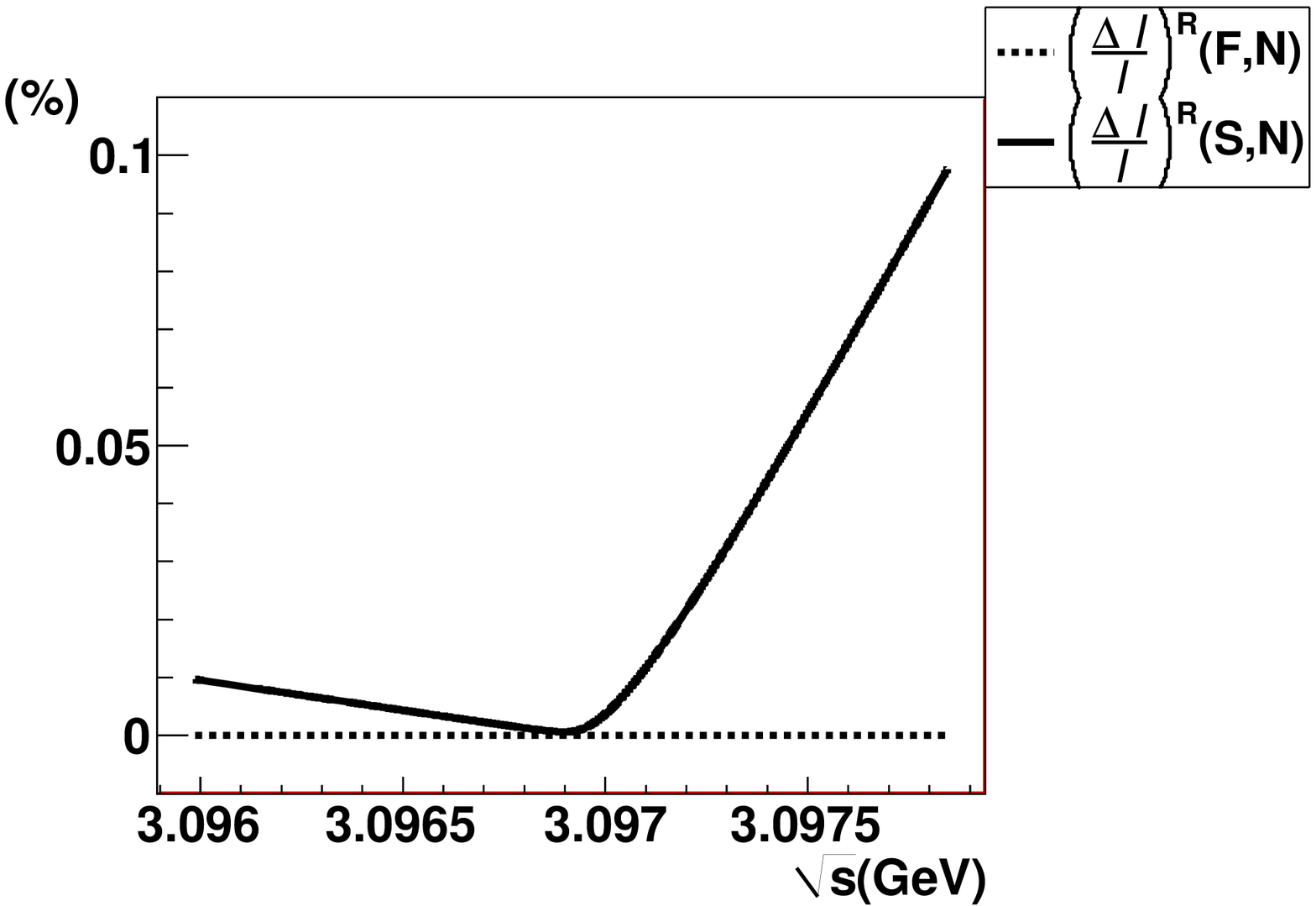}}
		\subfigure{\includegraphics[width=0.4\textwidth]{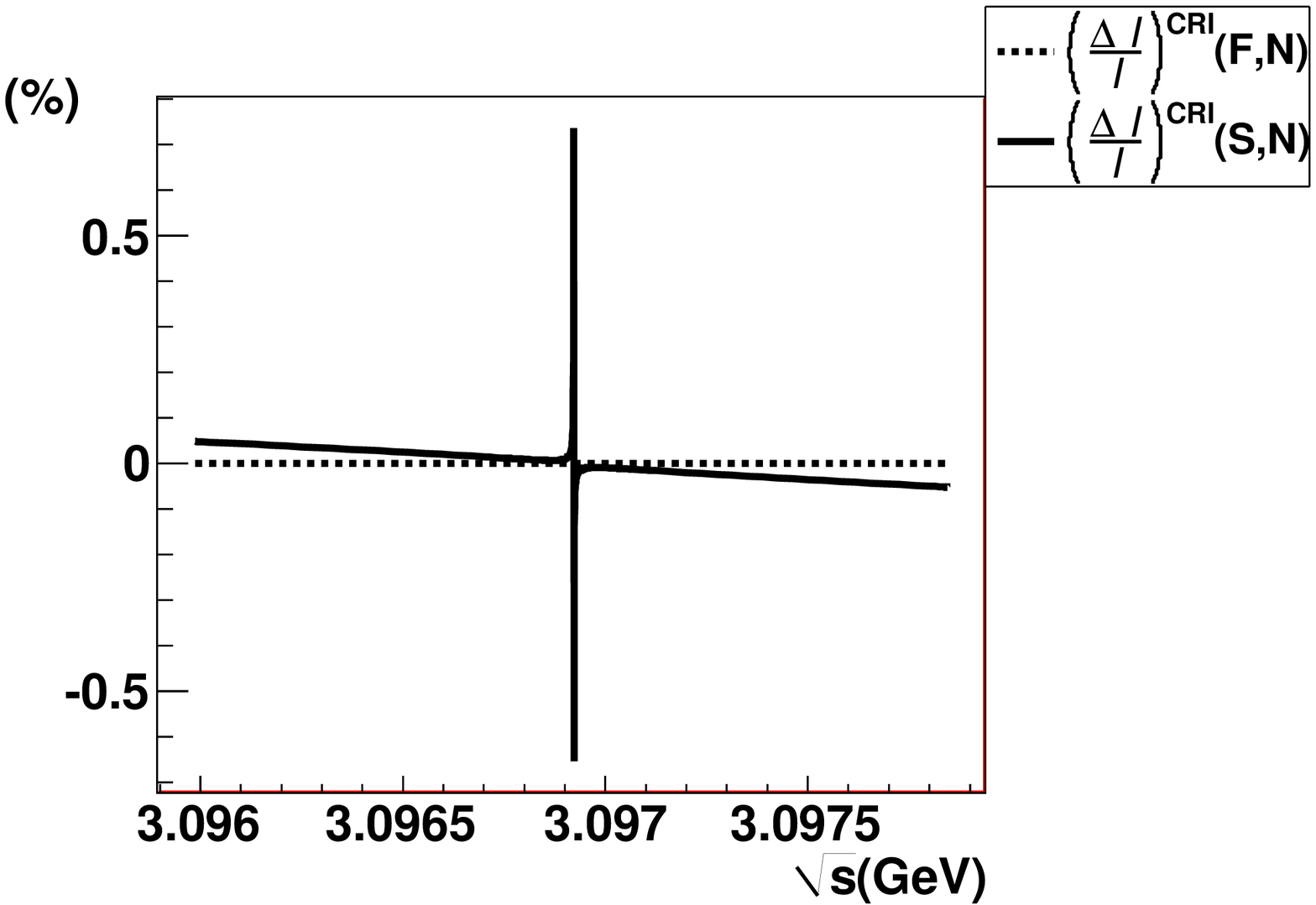}}
		\caption{Comparisons of analytic formulae with numerical computing results. In the middle of the right-hand plot, the dotted line has a similar structure to the solid one. It does not show clearly in the plot because of its small scale.}
		\label{Figure: Comparisons with numerical computing results.}
	\end{figure}
	\begin{multicols}{2}
		The variables in the legends are defined as
		\begin{equation*}
			\left(\frac{\Delta I}{I}\right)^{\rm R|CRI}(\text{F}|\text{S},\text{N}) = \frac{I^{\rm R|CRI}(\text{F}|\text{S})-I^{\rm R|CRI}(\text{N})}{I^{\rm R|CRI}(\text{N})}.
		\end{equation*}
		Here, the symbols $|$, F, S and N are same as those used at the beginning of Subsections \ref{Subsection: Applications of the structure fuction method to eetoee and eetomumu} and \ref{Subsection: Comparisons with numerical computing results}.
		
		As can be seen from the dotted lines, the full versions of the analytic formulae agree very well with the numerical computing results. In fact, detailed numbers show that their relative differences are less than 0.01\%. Similarly, from the solid lines, one can see that except for $I^{\rm CRI}$ at energies very close to the \jpsi peak, the simplified versions of the analytic formulae agree with the numerical computing results to better than 0.1\%. The upward and downward peaks of $\left(\frac{\Delta I}{I}\right)^{\rm CRI}(\text{S},\text{N})$ at energies near the \jpsi peak is caused by the smallness of the absolute values (very close to 0) of $I^{\rm CRI}$, which makes $\sigma^{\rm CRI}$ values negligible when compared with their corresponding $\sigma^{\rm R}$ values. Because in the end, only the sum of $\sigma^{\rm R}$ and $\sigma^{\rm CRI}$ will be used in our data analysis, the peaks of $\left(\frac{\Delta I}{I}\right)^{\rm CRI}(\text{S},\text{N})$ are not worrying for us.
		
	\end{multicols}
	
\end{small}


\begin{multicols}{2}
	
\end{multicols}

\clearpage
\end{CJK*}
\end{document}